# Plant and insect proteins support optimal bone growth and development; Evidences from a pre-clinical model

Gal Becker, Jerome Nicolas Janssen, Rotem Kalev-Altman, Dana Meilich, Astar Shitrit, Svetlana Penn, Ram Reifen and Efrat Monsonego Ornan*

Institute of Biochemistry and Nutrition, The Robert H. Smith Faculty of Agriculture, Food and Environment, The Hebrew University of Jerusalem, Rehovot 7610001, Israel

* Correspondence: efrat.mo@mail.huji.ac.il

**Abstract**

By 2050, the global population will exceed 9 billion, demanding a 70% increase in food production. Animal proteins alone may not suffice and contribute to global warming. Alternative proteins such as legumes, algae, and insects are being explored, but their health impacts are largely unknown. For this, three-week-old rats were fed diets containing 20% protein from various sources for six weeks. A casein-based control diet was compared to soy isolate, spirulina powder, chickpea isolate, chickpea flour, and fly larvae powder. Except for spirulina, alternative protein groups showed comparable growth patterns to the casein group. Morphological and mechanical tests of femur bones matched growth patterns. Caecal 16S analysis highlighted the impact on gut microbiota diversity. Chickpea flour showed significantly lower α-diversity compared with casein and chickpea isolate groups while chickpea flour, had the greatest distinction in β-diversity. Alternative protein sources supported optimal growth, but quality and health implications require further exploration.

## Introduction

The world's population is continuously growing and is projected to reach 9.7 billion by 2050[1]. Additionally, as developing countries become urbanized and income increases, they can afford higher-priced food items such as meat, leading to transformations in food consumption patterns. The increasing consumption of the current conventional protein sources, such as meat, dairy, fish, and eggs, leads to higher greenhouse gas (GHG) emissions. Moreover, the increased demand and thus the production of more animal feed leads to the conversion of forests and natural grasslands into agricultural lands[2]. By 2050 these dietary trends, will be a major contributor to global land clearing and an estimated 80% increase in global agricultural GHG emissions from food production[3]. There is growing interest in shifting towards a more plant-based diet due to its environmental sustainability compared with animal food production[4]. Finding sustainable and healthy protein sources is essential. Legumes, algae, and insects have emerged as promising alternatives.

Legumes like soy and chickpea, have low GHG and water footprints and enrich soil through nitrogen fixation. Soy protein, containing 36.3-47% total protein, with a Protein

Digestibility Corrected Amino Acid Score (PDCAAS) of 1.00, is high-quality and comparable to some animal-sourced proteins[5]. Chickpea protein, 17–22%, can be dry- or wet-extracted[6], with extruded chickpeas having the highest PDCAAS score (0.84)[7]. Chickpea also provide isoflavones, vitamins, minerals, antioxidants and healthy fatty acids[8].

Insects are consumed by 2 billion people and promoted by the FAO for their high protein, nutrient-dense fats, and minerals. They are environmentally friendly, emitting fewer GHGs and requiring no land clearing[9]. This study examines reduced-fat protein powder from the Mediterranean fruit fly.

Spirulina algae are single-cell blue-green organisms rich in protein (60-70%) and omega-3 fatty acids[10]. They also contain phycocyanin, an antioxidative pigment with anti-inflammatory and neuroprotective effects[11].

Proteins are essential for the body's structural and functional elements, exchanging nitrogen with the environment. Adequate high-quality protein intake is crucial for development and health, especially during growth, pregnancy, and lactation[12]. The skeleton is an ideal model to study dietary protein intake and amino acid supply. Nutritional deficiencies can lead to shorter, low-quality bones by affecting bone matrix protein synthesis and IGF1 expression, which regulates bone growth and calcium absorption[13]. Diet impacts bone health also through the gut microbiota (GM), which is influenced by dietary components. Over 50% of gut microbiota variations are diet-related, suggesting dietary strategies for disease management through GM modulation[14,15].

This study aims to determine which alternative proteins can substitute everyday proteins, considering growth, bone quality, and the gut microbiome impact.

## Results

### The effect of alternative proteins on growth parameters

We used a pre-clinical model to understand the health consequences of alternative protein consumption. A 6-week experiment was conducted on 3-week-old SD rats to evaluate growth, in terms of body weight (g) and length (cm), and particularly skeletal development, using different protein sources. This period mimics post-weaning human growth up to sexual maturity[16]. Weight was measured twice a week and body length, from nose to tail, was measured weekly to indicate longitudinal bone growth. The casein-based diet (Cas) served as the control group for normal growth patterns[17], as it is the recommended diet for rodents. Despite identical nutrient ratios, growth patterns varied across diets. The spirulina-protein-based diet (Spl) resulted in the slowest growth compared to Cas and other proteins (Fig. 1 A-C)[17]. After six weeks, significant differences were observed. The Spl group's mean weight (165.5g) was significantly lower than Cas (198.6g) and all other groups. The Fly diet group reached the highest final weight (236.32g), similar to the Soy, CP/I, and CP/F groups, which did not differ from the Cas group (Fig. 1 A, C). In terms of longitudinal growth, the Cas, Soy, CP/I, CP/F, and Fly groups showed similar body and femur lengths, while the Spl group had significant growth inhibition in both (Fig. 1 B-D).

During the growth period, both elongation and fat accumulation can affect the weight. To better define growth status, a weight-to-length index was calculated. The hierarchy between the groups remained: Soy, CP/I, and CP/F groups were similar Cas group. The Fly group had the highest weight-to-length index, indicating fat accumulation, while the Spl group had the lowest, suggesting growth inhibition (Fig 1E).

**Diet efficiency ratio of different proteins**

Throughout the experiment, food consumption was monitored bi-weekly to assess energy and protein intake by rats. Results show that all alternative protein groups consumed more food compared to the Cas group, both in grams and calories. Among them, the Fly group exhibited the highest food consumption (Fig. 1F). The energy efficiency ratio (EER), representing total weight gain per 100 kcal intake, was calculated. Most groups showed similar EER values, except for the Spl group, which demonstrated a lower EER. This finding suggests that despite consuming as much calories as other groups, the Spl group had lower protein efficiency and did not achieve expected growth (Fig. 1G).

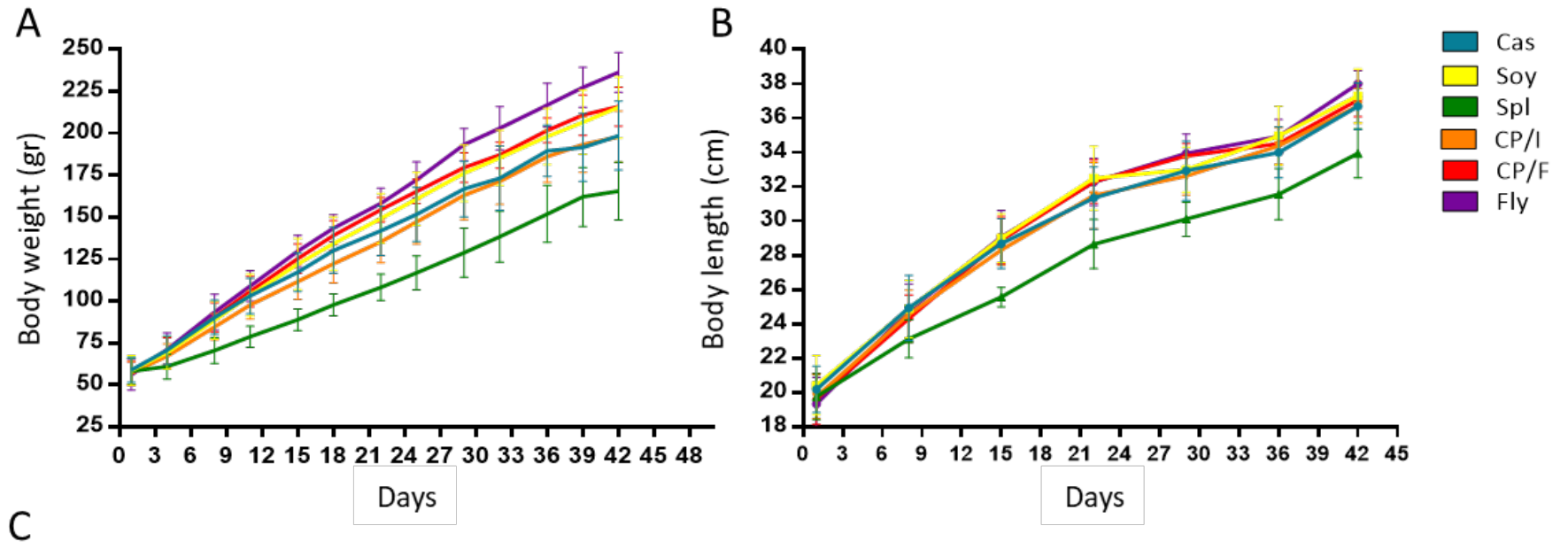

C

| Day | Cas | Soy | Spl | CP/I | CP/F | Fly |
|---|---|---|---|---|---|---|
| **Body Weight** | | | | | | |
| 1,4 | a | a | a | a | a | a |
| 8 | a | ab | c | abc | a | a |
| 11 | a | a | b | a | a | a |
| 15-22 | ab | ab | c | b | ab | a |
| 25-42 | b | ab | c | b | ab | a |
| **Body Length** | | | | | | |
| 1,8 | a | a | a | a | a | a |
| 15 | a | a | c | ab | a | a |
| 22 | ab | a | c | ab | a | a |
| 29-42 | a | a | b | a | a | a |

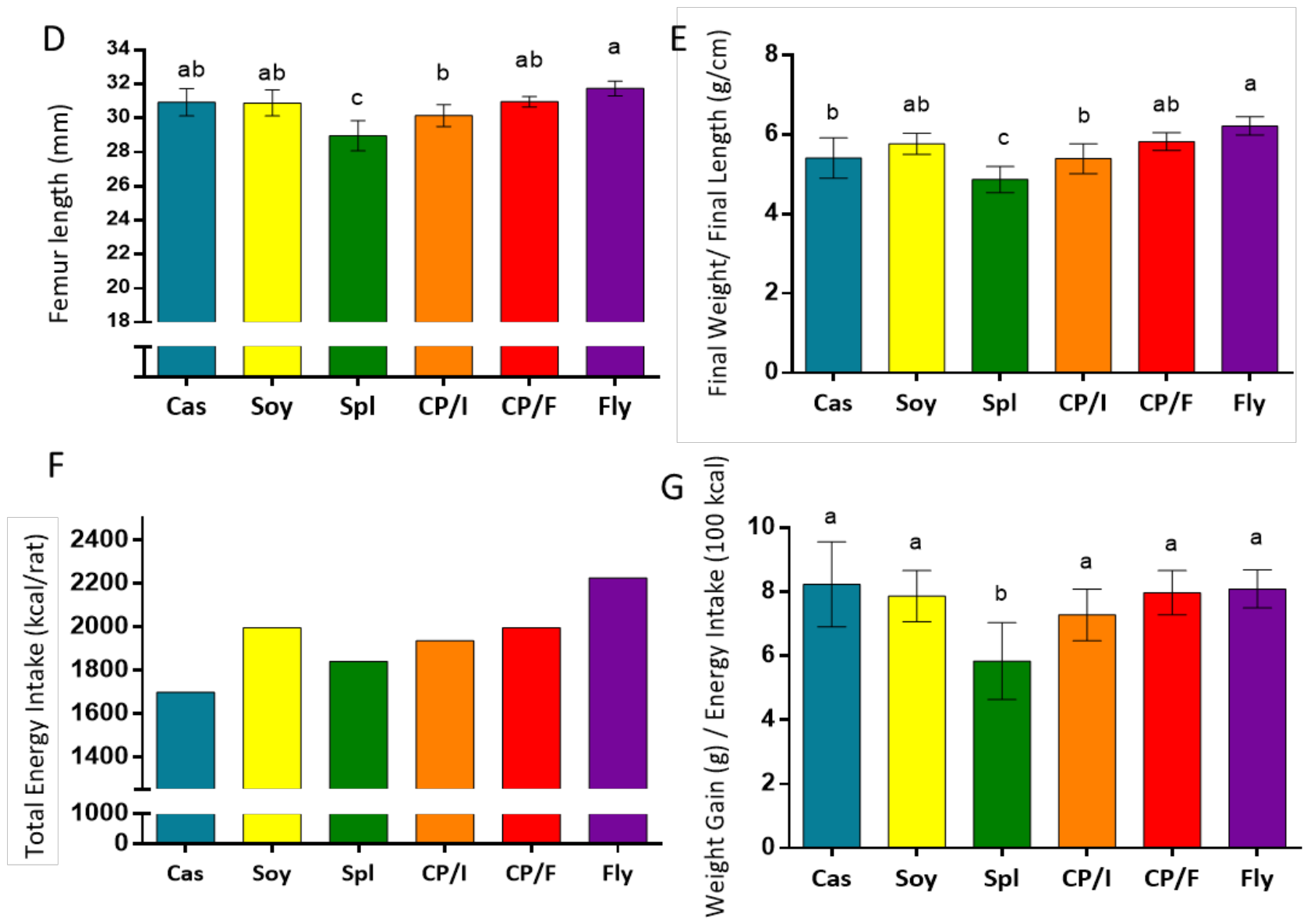

**Figure 1. Growth patterns of the groups and energy consumption**. (A) Body weight (g) and (B) body length (cm) throughout the experiment. (C) Table of statistical analysis (Graphs 1A-B) (D) Right femur length was measured by micro-CT (mm) at the age of 9 weeks. (E) Final Weight to Final length (g/cm). (F) Total energy intake (kcal/rat). Values are calculated as the mean of 2 cages for each diet and expressed as the estimated consumption for rats. (G) Weight gain/energy intake (g/100 kcal). Statistical differences in body weight and length are presented according to measurements days. Values are expressed as mean ± SD of n=8 rats/group, different superscript letters are significantly different ($P < 0.05$) by one-way ANOVA followed by Tukey's test.

## The effect of alternative protein on bone morphological and mechanical parameters

Next, we examined whether different protein sources affected bone quality during the fast-growing period, in addition to their effects on growth patterns[18]. Bone mechanical performance is crucial for functional load-bearing, with fractures occurring when this fails. The key structural parameters influencing bone performance are geometry, mineral density, and cortical bone integrity[19]. Using Micro-CT, we analyzed the proximal to mid-diaphysis region of the femora from all rats. A three-point bending experiment evaluated the mechanical properties of the bones (Table 1, Fig. 2)[18,20].

A comparison of bones from the various groups showed minor differences in femur trabecular parameters. Trabecular number and separation did not differ. However, BV/TV, which represents the trabecular bone volume out of the total volume, in the Spl group (36.42%) was statistically lower than in CP/F and Fly (44.42%, 45.41%) due to reduced trabecular thickness (Table 1). Cortical bone analysis revealed that alternative protein (Soy, CP/I, CP/F, Fly) did not differ from the casein group in cortical area Ct. Ar, the ratio between Ct.Ar to total area (Ct.Ar/ Tt.Ar), cortical thickness(Ct. Th), medullary area (Ma. Ar) and bone mineral density (BMD). However, the Spl group had statistically lower Tt.Ar, Ct.Ar, Ct.Ar/Tt.Ar, and Ct.Th, compared to the casein group. BMD values were regular across all diets with no statistical differences (Table 1).

**Table 1. Morphometric characteristics of trabecular and cortical bone of the femur**. Bone volume over total volume, BV/TV; Trabecular number, Tb.N; Trabecular separation, Tb.Sp; Trabecular thickness, Tb.Th, Total Cross-Sectional Area, Tt.Ar; Cortical Bone Area, Ct.Ar; Cortical area fraction, Ct.Ar/Tt.Ar; Cortical thickness, Ct.Th; Medullary area, Ma.Ar; Bone mineral density, BMD. Values are expressed as mean ± SD of n=8 rats/group, different superscript letters are significantly different ($P < 0.05$) by one-way ANOVA followed by Tukey's test.

| Trabecular parameter | Cas | Soy | Spl | CP/I | CP/F | Fly |
|---|---|---|---|---|---|---|
| **BV/TV (%)** | $38.7 \pm 5^{ab}$ | $41.8 \pm 5^{ab}$ | $36.4 \pm 5.6^{b}$ | $40.6 \pm 3.6^{ab}$ | $44.4 \pm 2.8^{a}$ | $45.4 \pm 3.1^{a}$ |
| **Tb.N** | $3.4 \pm 0.3^{a}$ | $3.41 \pm 0.3^{a}$ | $3.26 \pm 0.2^{a}$ | $3.42 \pm 0.2^{a}$ | $3.44 \pm 0.2^{a}$ | $3.54 \pm 0.2^{a}$ |
| **Tb.Sp** | $0.23 \pm 0.05^{a}$ | $0.22 \pm 0.03^{a}$ | $0.25 \pm 0.03^{a}$ | $0.22 \pm 0.03^{a}$ | $0.21 \pm 0.02^{a}$ | $0.22 \pm 0.05^{a}$ |
| **Tb.Th** | $0.11 \pm 0.005^{b}$ | $0.12 \pm 0.007^{ab}$ | $0.11 \pm 0.013^{b}$ | $0.11 \pm 0.007^{ab}$ | $0.13 \pm 0.05^{a}$ | $0.13 \pm 0.005^{a}$ |
| **Cortical parameter** | | | | | | |
| **Tt.Ar ($mm^2$)** | $9.51 \pm 0.9^{a}$ | $9.03 \pm 0.7^{ab}$ | $7.6 \pm 0.6^{c}$ | $8.53 \pm 0.6^{bc}$ | $9.28 \pm 0.3^{ab}$ | $9.96 \pm 0.6^{a}$ |
| **Ct.Ar ($mm^2$)** | $4.62 \pm 0.3^{ab}$ | $4.59 \pm 0.3^{ab}$ | $3.5 \pm 0.2^{c}$ | $4.4 \pm 0.2^{b}$ | $4.69 \pm 0.1^{ab}$ | $4.93 \pm 0.2^{a}$ |
| **Ct.Ar/Tt.Ar (%)** | 48.7±2.9$^{ab}$ | 50.9±2.1$^{a}$ | 46.2±2.7$^{b}$ | 51.7±2.8$^{a}$ | 50.5±1.6$^{a}$ | 49.6±2$^{ab}$ |
| **Ct.Th (mm)** | 0.45±0.02$^{a}$ | 0.47±0.03$^{a}$ | 0.38±0.02$^{b}$ | 0.46±0.03$^{a}$ | 0.47±0.02$^{a}$ | 0.47±0.01$^{a}$ |
| **Ma.Ar (mm$^2$)** | 4.88±0.7$^{ab}$ | 4.44±0.4$^{ab}$ | 4.1±0.5$^{b}$ | 4.13±0.5$^{b}$ | 4.59±0.3$^{ab}$ | 5.03±0.5$^{a}$ |
| **BMD (g/cm$^3$)** | 1.04±0.12$^{a}$ | 1.07±0.12$^{a}$ | 0.93±0.18$^{a}$ | 0.99±0.16$^{a}$ | 0.99±0.1$^{a}$ | 0.94±0.09$^{a}$ |

The biomechanical properties of the femur, studied by the three-point bending experiment[18], revealed no significant differences between all proteins except for the Spl group, as demonstrated by the force-displacement curves (Fig. 2). The Spl group exhibited inferior mechanical results for yield load, fracture load, and maximal load parameters compared with the Cas group. Thus, in terms of femur morphological, structural, and mechanical parameters, we did not see a deterioration when the protein source was replaced from casein to the alternative proteins excluding spirulina.

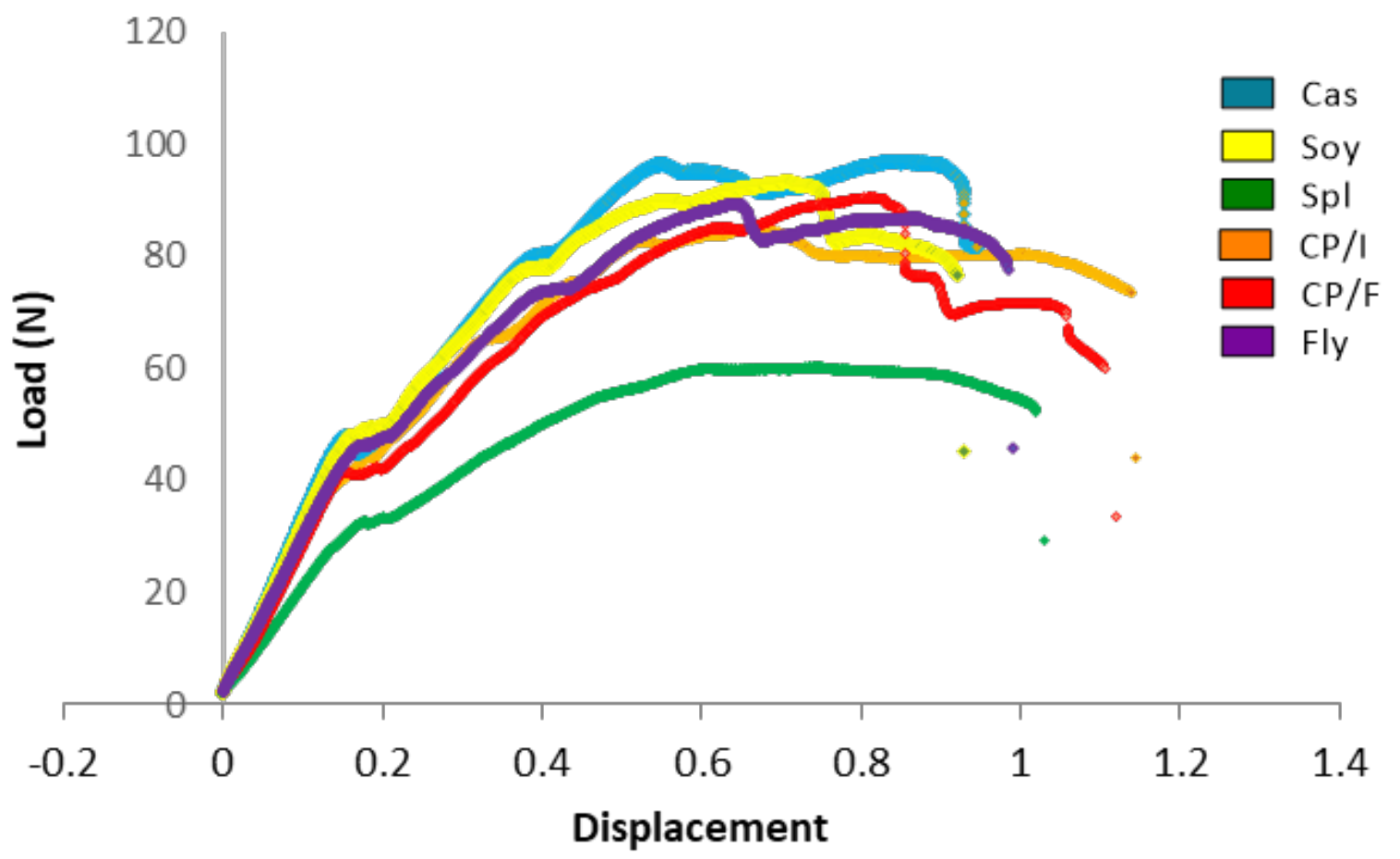

| The tested parameter | Cas | Soy | Spl | CP/I | CP/F | Fly |
|---|---|---|---|---|---|---|
| Slope (N/mm) | 285.2±46.7$^{a}$ | 294.8±36.3$^{a}$ | 173.7±37$^{b}$ | 285.1±29$^{a}$ | 275.2±34.9$^{a}$ | 254.2±61.1$^{a}$ |
| Young's modulus (N/mm$^2$) | 2367.7±416.7$^{a}$ | 2396.7±36.3$^{a}$ | 2102.8±37$^{b}$ | 2689.3±29$^{a}$ | 2135±34.9$^{a}$ | 2054.6±61.1$^{a}$ |
| Yield load (N) | 40±3.9$^{ab}$ | 48.6±4.4$^{a}$ | 31.5±1.8$^{b}$ | 44.3±4.4$^{a}$ | 44.8±3.9$^{a}$ | 43.1±3.8$^{ab}$ |
| Max load (N) | 86.4±7.6$^{a}$ | 92.1±7.3$^{a}$ | 61.3±6.7$^{b}$ | 87.4±5.7$^{a}$ | 86.4±3.3$^{a}$ | 90±6.5$^{a}$ |
| Fracture load (N) | 73.7±10.4$^{a}$ | 74.7±16.9$^{a}$ | 40.5±10.8$^{b}$ | 70.1±14.9$^{a}$ | 65.7±6.2$^{a}$ | 71.2±10.4$^{a}$ |
| Energy to fracture (N×mm) | 49.8±18.9$^{a}$ | 49.1±15.7$^{a}$ | 58.5±12.6$^{a}$ | 49±14.4$^{a}$ | 53.9±3.7$^{a}$ | 58±9.8$^{a}$ |

**Figure 2. Force-Displacement curves of experimental groups**. Mechanical properties were evaluated using a three-point bending experiment performed on the bones of all rats from the six groups, representative results are shown. Values are expressed as mean ± SD of n=8 rats/group, different superscript letters are significantly different ($P < 0.05$) by one-way ANOVA followed by Tukey's test.

**Possible mechanisms underlying the effect of alternative proteins on growth**

Next, we explored the impact of different dietary protein sources on growth and skeletal quality by analyzing amino acid composition and microbiome modifications.

### Amino acid composition

Humans and rats require nine essential amino acids through diet. Protein quality is indicated by the ratio of the “most limiting amino acid” to its requirement, reflecting efficiency in utilization and protein synthesis [21]. The amino acid composition was analyzed by ion-exchange chromatography and compared to the requirements of growing rats. Tryptophan and the sulfur containing amino acids (methionine and cysteine) in casein are the most limiting amino acids, composing about 55%, and 69% of rats’ nutritional requirement, respectively (Table S1).

Due to food consumption variations among groups, we further analyzed amino acid intake. Comparing other diets to casein, as the recommended protein source[17], table 2 showed that methionine and cysteine were most limiting in the Soy and CP/I diets, valine in the CP/F diet, and all amino acids were higher in the Fly diet. The Spl diet had significant lysine deficiency, with the spirulina group consuming nearly half the lysine compared to the casein group, possibly explaining growth inhibition.

**Table 2. Essential amino acid consumption throughout the experiment.** Amino acids intake from the diet by the 6 experimental groups (g/rat). The amount eaten by each group multiplied by the AA profile of protein in the diet. The Cas group was used as a reference (100%).

| AA eaten (g/rat) | Cas | Soy | Spl | CP/I | CP/F | Fly |
|---|---|---|---|---|---|---|
| **Histidine** | 2.22(100%) | 3.11(140%) | 1.65(74%) | 2.34(106%) | 3.25(146%) | 3.35(151%) |
| **Isoleucine** | 3.6(100%) | 4.73(131%) | 4.33(120%) | 3.85(107%) | 4.1(114%) | 5.97(166%) |
| **Leucine** | 7.72(100%) | 9.07(118%) | 7.10(120%) | 7.51(97%) | 8.35(108%) | 10.5(136%) |
| **Lysine** | 8.03(100%) | 7.71(96%) | 4.76(59%) | 7.37(92%) | 8.35(104%) | 10.27(128%) |
| **Methionine+Cystein** | 2.91(100%) | 2.73(94%) | 3.02(104%) | 2.07(71%) | 3.19(109%) | 3.66(126%) |
| **Phenylalanine+Tyrosine** | 9.09(100%) | 10.87(120%) | 7.57(83%) | 8.81(97%) | 10.09(111%) | 15.41(170%) |
| **Threonine** | 3.68(100%) | 5.06(137%) | 4.50(122%) | 3.25(88%) | 5.04(137%) | 6.31 (171%) |
| **Valine** | 4.93(100%) | 5.32(108%) | 4.21(85%) | 4.03(82%) | 4.73(96%) | 6.59(134%) |
| **Tryptophan** | 0.74(100%) | 1.24(169%) | 1.03(141%) | 0.57(77%) | 1.35(184%) | 1.5(205%) |

### Microbiome analysis

The microbiome appreciably influences bone homeostasis and disease[22]. To understand its role in our model, we analyzed rat’s cecum samples for microbiome composition and diversity. 16S sequencing covered ~38,000 reads per sample, ranging from ~15,000 to ~58,000 reads. We unambiguously detected 11 phyla, 55 families, and 93 genera across all samples. Figure S3 shows the abundance and prevalence of the most abundant genera. The highest abundance was detected for several *Lactobacillaceae* (*Lactobacillus*, *Ligilactobacillus*, *Limosilactobacillus*) and *Ruminococcaceae* (*Ruminococcus*, *Flintibacter*, *Lawsonibacter*). Conversely, *Butyricicoccus* was present only in a subset of samples, indicating a strong correlation with a specific diet.

The α-diversity of all groups, measured using the Shannon index, revealed a significantly lower index in CP/F, while other groups displayed a higher index compared to the Cas group (Fig. 3A). A significant difference in α-diversity was also found between the CP/I and CP/F groups, highlighting the impact of chickpea processing (protein isolation) on the microbiome.

β-diversity is illustrated in the NMDS plot of the Bray-Curtis dissimilarity (Fig. 3B). Samples from all groups clustered together, but only the Soy and CP/I groups showed overlapping clusters (Fig. 3B). Differences in microbiome composition between groups were tested first for homogeneity of sample dispersion, which was not met (p.adj < 0.05) for the comparisons between the CP/I group with the Cas, Spl and Fly groups, as well as between the Soy group and the Cas and Spl groups. PERMANOVA analysis of Bray-Curtis dissimilarities revealed significant differences between all groups (p.adj < 0.05), indicating distinct microbiome compositions only for groups with similar sample dispersion. Interestingly, β-diversity significantly differed between the CP/I and CP/F groups. Moreover, the high-fiber Spl and CP/F diet showed distant clustering from other groups. The highly processed Cas, Soy, and CP/I samples (91%, 90%, 89% protein in the product) clustered together, while the less processed, natural fiber-rich, CP/F, and Spl groups (20%, 60% protein in the product) clustered distantly. The intermediate processed Fly diet (66%), clustered near the highly processed diets.

Differential abundance analysis between Cas and all other groups showed 11 genera in Soy, 22 in Spl, 9 in CP/I, 21 in CP/F, and 12 in Fly differing from the Cas group (Fig. 3C and Table S2). The heatmap was divided into 4 clusters: The first cluster contained *Faecalibaculum* that was less abundant in all groups compared to Cas. *Clostridium sensu stricto* was less abundant in all groups except for CP/I compared to Cas, whereas *Bifidobacterium* was less abundant in all groups except for CP/F compared to Cas. *Lactobacillus* was less abundant Soy, CP/I and Fly compared to the Cas group.

The second cluster is composed of genera being less abundant in the CP/F group compared to Cas, such as *Lawsonibacter*, *Desulfovibrio* and *Romboutsia*. *Staphylococcus* was more abundant in Fly and Soy compared to Cas. *Streptococcus* was more abundant in Fly and Spl compared to the Cas group. *Clostridium XVIII* was more abundant in the Fly group but less abundant in all other groups compared to Cas. *Flintibacter* was less abundant in the Spl compared to the Cas group. *Clostridium IV* was more abundant in CP/I and *Duncaniella* more abundant in the Soy, Spl, CP/I and CP/F groups compared to the Cas group.

The third cluster had *Coprobacillus*, *Erysipelatoclostridium*, *Roseimaritima*, *Rubinisphaera,* and *Saliniramus* more abundant in the Spl compared to the Cas group. *Parabacteroides* was more abundant in Spl, CP/F and Fly compared to the Cas group, whereas *Mucispirillum* was less abundant in the CP/F compared to the Cas group.

The fourth cluster contained bacteria with a higher abundance in the CP/F, including *Butyricicoccus* and *Parasutterella* groups as compared to Cas. *Bacteroides* had a higher abundance in the CP/F but a lower abundance in the Soy and CP/I compared to the Cas group.

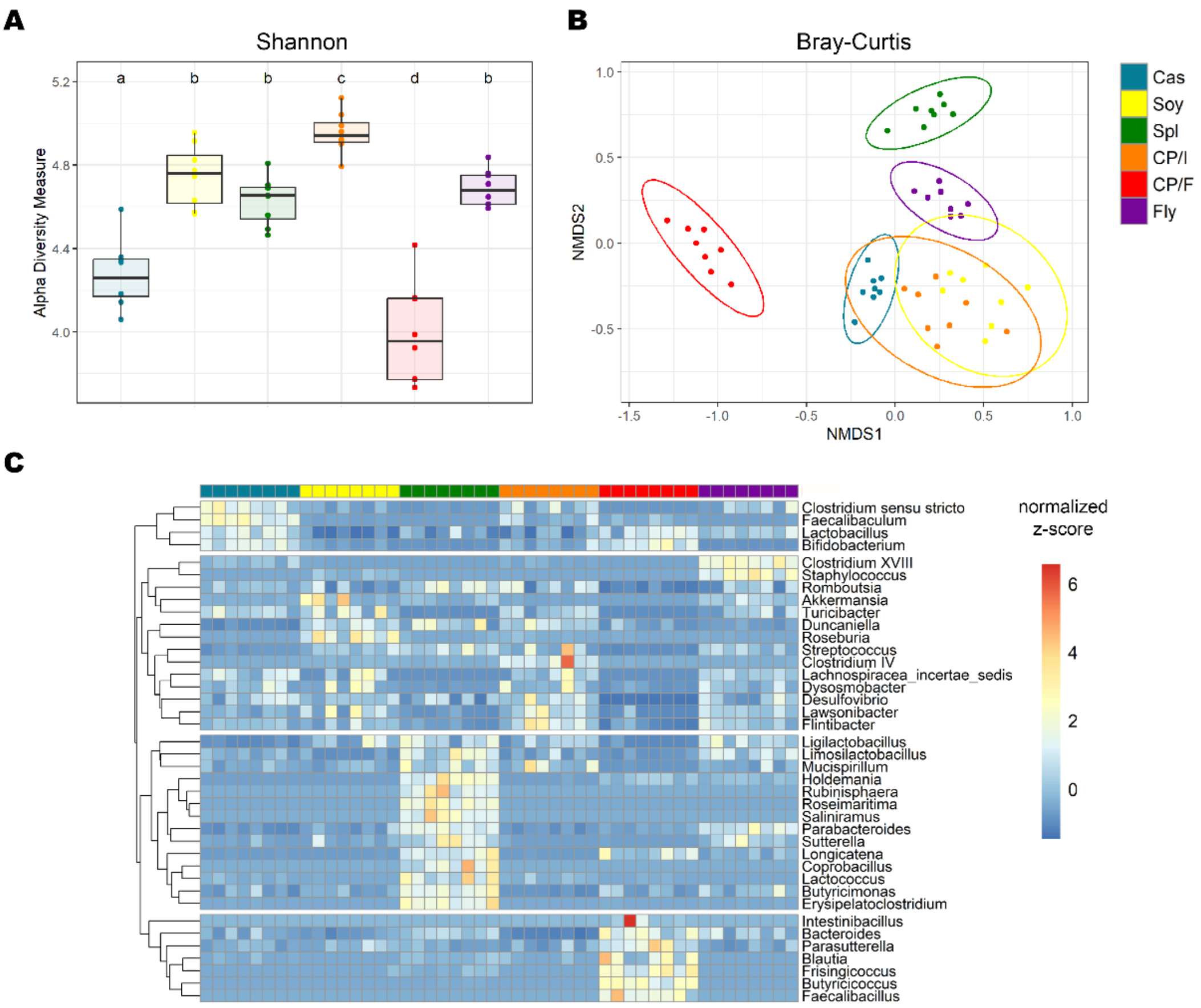

***Figure 3. Effect of alternative protein consumption on the gut microbiota.*** **(A)** Alpha diversity was calculated using the Shannon index and a pairwise Wilcoxon rank sum test with FDR adjustment. **(B)** NMDS plot of Bray-Curtis dissimilarity. Plot ellipses represent the 95% confidence regions for group clusters. PERMANOVA showed significant differences between all groups. However, homogeneity of dispersion was not given for the comparisons of Soy vs Cas, Soy vs Spl, CP/I vs Cas, CP/I vs SPL, and CP/I vs Fly. *P* values were adjusted using Holm's method. **(C)** Differential abundance of bacterial genera in alternative protein groups compared to the Cas group, analyzed using DESeq2 with FDR p.adj < 0.05, log2FC ≥ 1, and sum of counts across all samples ≥ 50. Bacteria not classified on genus level were excluded. Colors indicate row-wise z-score of relative genera abundance.

## Food processing affects the microbiome composition

Focusing on differences in diet protein processing (isolated vs. non isolated sources), we found an altered abundance of *Parabacteroides* and *Turicibacter* in the less processed Spl, CP/F, and Fly compared to the Cas group, with no significant difference in the Soy and CP/I compared to the Cas group, all containing higher processed proteins. To further investigate how the microbiome is affected by processed diet protein, we performed differential abundance analysis between the CP/I and CP/F groups. 13 families and 35 genera were explicitly classified on the respective taxonomic level (Fig. 4, Tab. S3 and Tab. S4). *Bifidobacterium* and *Bilophila* (*Bifidobacteriaceae* family), *Lachnospiraceae* and *Lactobacillaceae* families and their members, *Blautia* and *Lactobacillus* were more abundant in CP/F group. However, other members of the respective family, including *Roseburia* and *Ligilactobacillus* were less abundant in the CP/F group. Finally, a higher number of genera was found to be more abundant in CP/I, e.g., *Paramuribaculum*, *Butyricimonas*, *Eubacterium,* and *Desulfovibrio*.

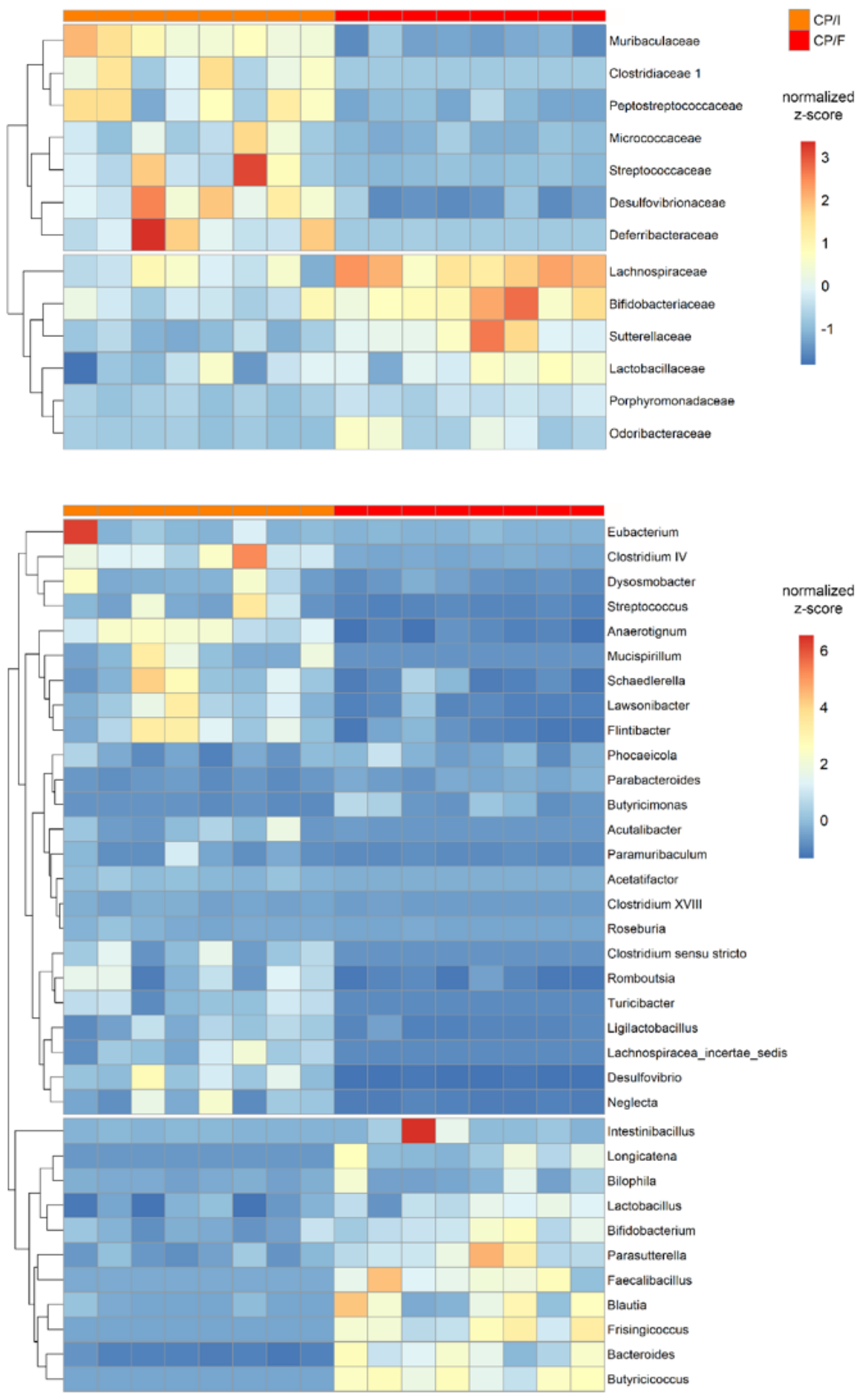

**Figure 4: Differential abundant bacteria between the more strongly processed CP/I diet protein group and the CP/F group.** Differential bacterial abundance on (top) family and (bottom) genus level comparing the CP/I and CP/F groups, analyzed using DESeq2 with FDR p.adj < 0.05, log2FC ≥ 1, and sum of counts across all samples of all experimental groups ≥ 50. Bacteria not classified on family or respectively genus level were excluded. Colors indicate row-wise z-score of relative family or respectively genera abundance.

## Discussion

This study investigates the impact of several protein sources on skeletal development and bone quality, considering the growing concerns about environmental sustainability, public health, and ethics. Efforts to find alternatives to animal-based protein are increasing[23]. Before widespread adoption, it's crucial to evaluate these protein's health effects during development. Our findings revealed that soy, chickpea (isolated and flour), and fly larvae proteins supported growth and bone quality comparably to casein protein. However, spirulina protein was less effective in promoting growth and bone health, possibly due to its lysine deficient and influence on gut microbiota.

**Protein quality evaluation through sensitive model - beyond PDCAAS and DIAAS**

When substituting animal-based proteins with alternatives, amino acid composition is crucial. The gold standard for growth and development is casein, a milk protein[17,24]. Plant-derived proteins are considered nutritionally inadequate because cereals lack lysine, and legumes are low in methionine and cysteine[25]. However, our experiment found that most tested protein sources, supported growth parameters comparable to casein, despite conventional evaluation methods assigning low scores to most plant proteins. Spirulina was the only protein source causing growth delay, attributed to diet composition rather than lack of calories. Notably, our results did not correlate with the commonly used protein quality scores, PDCAAS and DIAAS[26,27,28,29], highlighting their limitations. For instance, extruded chickpea has a PDCAAS of 0.83 and baked chickpea has a DIAAS of 0.84, compared to casein's PDCAAS of 1.00 and DIAAS of 1.08 [7,30]. Yet, chickpea flour and isolate in our study matched casein growth performance. Conversely, despite a DIAAS of 0.85 for spirulina[31], it showed the lowest growth bone and quality parameters in our study.

DIAAS and PDCASS scores for protein quality overlook variations in protein quantities and interactions with other dietary components. Furthermore, plant-based protein sources often contain anti-nutritional factors such as trypsin inhibitors and tannins[32]. These elements can impede protein digestion, potentially increasing nitrogen losses through feces, which lower apparent protein digestibility and may elevate protein requirements[33,34]. PDCAAS does not account for these effects, leading to an overestimation of protein quality in items containing these natural anti-nutritional factors[35].

Significant discrepancies exist in DIAAS and PDCAAS values for the same protein sources across various studies[28,36]. For instance, reported DIAAS values for soy products can range widely, from 0.75 to 0.45[37]. To address these inconsistencies, we propose a novel approach using an *in-vivo* physiological preclinical model[38,39]. This model allows for rapid and comprehensive analysis of not only pure proteins but also of the effects of processing methods, protein quantities in the diet, interactions with other nutrients, and combinations of diverse protein sources.

**Exploring the mechanism behind spirulina's growth retardation effect**

Spirulina, noted for its high nutritional value, is considered a potential alternative to traditional protein sources[40]. Due to compounds like phycocyanin, chlorophyll, carotenoids, and gamma-linolenic acid it also possesses antioxidant, anti-inflammatory, and bone-strengthening properties. In addition to promoting osteoblast activity and mineralization[41] Spirulina also contains essential minerals like calcium, magnesium, and zinc crucial for bone development[42,43]. Despite its benefits as a food supplement[44], our study found that using spirulina as the sole protein source during rats' growth resulted in reduced weight gain, slowed longitudinal growth, and modified bone strength parameters. Similar effects were observed in previous research where spirulina led to impaired growth, shorter femur, reduced mechanical strength, and lower femoral BMD[38].

In contrast, Fournier et al. (2016) showed that spirulina successfully replaced half of the casein in young rats fed a low-protein diet (10% protein) leading to improved bone biomechanical properties. Similarly, Cho et al. (2020) found that substituting casein with spirulina enhanced femur and lumbar spine length, increased bone strength and mineral content, and stimulated growth-regulating hormones[45]. One potential explanation for this discrepancy is the lower lysine content in the spirulina diet, compared to the casein used in Cho's study. Our analysis revealed that lysine, an essential amino acid crucial for growth, was the limiting AA in spirulina protein to 58% of the Cas group. Despite consuming more calories, the Spirulina group ingested 4.77 g of lysine per rat, while the Casein group consumed 8.03 g per rat. This supports previous research highlighting lysine's role in growth and development [46,47]. In addition, bacteria linked to bone diseases, such as *Erysipelatoclostridium* were more abundant in the Spl compared to the Cas group [48].

We suggest that spirulina alone may not adequately meet anabolic needs despite its benefits. However, studies indicated that using spirulina as a supplement to an insufficient diet or as a partial protein substitute can be beneficial[44,49]. Therefore, when replacing animal-based proteins with alternatives, exploring combinations of spirulina with other protein sources becomes crucial.

**The microbiome composition varied between all diets**

Microbiome dysbiosis, linked to skeletal health conditions[50], prompted us to conduct 16S sequencing of caecal samples. Common bacteria such as *Lactobacillus*, *Ruminococcus*, *Romboutsia*, *Clostridium sensu stricto*, *Turicibacter,* and *Akkermansia* were identified [51,52]. The Shannon index differed from the Casein group across all groups, but no consistent pattern regarding a correlation with bone quality emerged. The Spl group, which showed reduced measurements in length, weight, and the three-point bending assay, had a Shannon index between the Cas and non-phenotypic CP/I groups. Thus, relying solely on $\alpha$–diversity measurement is inadequate for distinguishing a healthy or pathogenic microbiome composition affecting skeletal growth[53]. The lower $\alpha$-diversity in CP/F compared to CP/I may stem from fiber-fermenting butyrate-producing bacteria outcompeting others in the caecum. $\beta$-diversity analysis revealed that diets based on protein isolates, including the Cas, Soy, and CP/I groups, cluster together, implying similar microbiome composition. This could be due to protein extraction methods influencing factors like pH, solubility, and protein folding, which in turn affect microbiome composition[54,55]. The Fly group clustered further away from other protein isolate groups, likely due to its lower processing grade. The Spl and CP/F groups' distant clustering was attributed to their lower protein processing and, high natural fiber content.

Differential abundance analysis revealed genera *Clostridium sensu stricto*, *Desulfovibrio*, *Duncaniella*, *and Faecalibaculum,* which are often abundant under pathologic conditions. Our previous study found an altered abundance of these bacteria in unbalanced diets[56]. However, here, these changes did not result in severe phenotypes. While these microbiome changes might be related to disease risk in adolescent and aging rats, this study focused on bone development and did not explore the potential long-term effects.

**Chickpeas as an example of the influence of food processing on the microbiome**

Using isolated proteins affects composition (fats, carbohydrates micronutrients, fibers) between chickpea isolate and chickpea flour-based diets. Higher *Paramuribaculum* abundance in a lower fiber diet was suggested to digest host glycans due to fewer indigestible nutrients in the large intestine[57], in agreement with our observation in the CP/I vs. CP/F groups. Dietary fibers promote fermentation, generating metabolites and protecting the intestine[58,59]. It has been reported that fibers increased the abundance of *Bifidobacterium*, *Lactobacillus, and Roseburia* in the caecum of rats[60], in agreement with our observations by comparing CP/I and CP/F. Among those *Roseburia* and *Lachnospiraceae* produce butyrate.

Nevertheless, other butyrate producers such as *Eubacterium*[61] or *Butyricimonas*[62] were found to be less abundant in CP/F. Finally, the abundance of other genera involved in butyrate processing was affected, e.g., the endotoxin-producing opportunistic pathogen *Desulfovibrio*, which damages the gut barrier and promotes inflammation by inhibiting butyrate oxidation and decreasing energy supply to colonocytes was less abundant in CP/F[63]. Overall, the microbiome composition of each diet group showed significant variations, possibly due to factors like diet processing or fiber content. Considering our bone and growth measurements, these differences (except for spirulina) lay in the natural spectrum of the rat microbiome and do not indicate a pathological shift or low protein digestibility.

**Modern food trend: Exploring Alternative Protein**

Interest in substituting animal-derived foods with alternatives has urged[64], due to ethical, environmental, and health reasons[65,66]. The rise of vegan and vegetarian diets, especially in Western cultures[67,68], is linked to longevity, improved quality of life, and reduced risk of non-communicable diseases[69,70]. However, vegan diets can lead to nutrient deficiencies[71,72], such as in vitamin B12, iron, calcium, and vitamin D[73,74]. For example, dietary vitamin B12 intake drops from 7.2 µg in meat-eaters to 0.4 µg in vegans[75], requiring careful planning and continuous monitoring[74]. Ultra-processed foods (UPF) in vegan diets is another concern, as it often include unhealthy components like saturated fats, starches, sugars, and additives[76]. UPF are formulated to mimic the physicochemical and sensory characteristics of animal-based products, but can adversely affect growth, and bone development and increase the risk of metabolic diseases, like obesity, diabetes, and coronary heart disease[39,77].

Shifting to plant-based diets can reduce the environmental footprint of animal agriculture, lowering greenhouse gas emissions, water usage, and deforestation[78]. Promoting locally sourced plant-based proteins can also enhance food security and create new markets, especially in animal farming regions[79].As global sustainability, health, and ethical concerns grow, finding suitable substitutes for traditional animal-based proteins becomes crucial. Our findings highlight the effects of these proteins during critical growth periods and their impact on gut microbiota and overall health

Understanding these interactions can help in developing comprehensive dietary recommendations supporting health, sustainability, and economic viability for future generations. The nutritional quality of food products varies due to factors like soil conditions, climate, storage, processing, and preparation[80,81], making nutritional investigation complex. This complexity is crucial for future generations as alternative proteins become more vital.

## Material and methods

### Animal experiment

Three-week-old Female Sprague Dawley (SD) rats after weaning, were purchased from Harlan Laboratories (Rehovot, Israel) and housed in environmentally controlled conditions. All procedures were approved by the Hebrew University Animal Care Committee (permit number AG-62804-01AG-). After 4 days of adaptation to the control chow diet, rats were randomly divided into 6 groups, 8 rats in each group. The control group consumed casein as a standard protein source; while the five remaining experimental groups consumed the same diet with an alternative protein source. All rats had *ad libitum* access to food and liquids. Throughout the experiment, body weight and food intake were measured twice a week. Food consumption was collected from 2 cages per experimental group, with 4 rats in each cage. The constellation chosen for the experiment, as required by ethics, does not enable the calculation of significance in food consumption. The other measurements (anthropometric, bone parameters, fat analyses, and microbiome) were done on each rat separately (n=8) thus allowing the statistical analyses. Energy efficiency was calculated by the added body weight of each rat divided by the average caloric intake. At the end of the experiment, after 6 weeks, the rats were anesthetized with isoflurane and femurs and were manually cleaned up of soft tissue, wrapped in saline-soaked gauze, and stored at -20°C until analysis (mechanical/ micro-CT testing). Tibias were fixed for histological analysis[26].

**Diet preparation and composition**

All groups consumed semi-purified diets,(Table 1): 18% fat, 61% carbohydrate, and 21% protein, according to the American Institute of Nutrition (AIN-93G) recommendations for rodents at growth phase[27]. The diets differed in their protein source, as follows: Standard control diet (Cas), protein from caseinate (MEGGEL); Soy diet (Soy) protein from soy isolate (MP biomedicals); Spirulina diet (Spl), protein from spirulina powder (Abundance); Chickpea isolate diet (CP/I), protein from chickpea isolate (ChickP); Chickpea flour diet (CP/F), protein from chickpea flour (Shay Shel Ha'teva); Fly diet (Fly), protein from fruit fly larvae powder (Flying Spark) (Table 1). All diets were supplemented with vitamins and minerals[27], the whole meal was homogenized, shaped as dumplings, and frozen at -20°C.

**Table 1. Materials and suppliers**

| | Cas | Soy | Spl | CP/I | CP/F | Fly |
|---|---|---|---|---|---|---|
| **Diet composition** | **g/997g** | **g/1001g** | **g/953g** | **g/992g** | **g/1086g** | **g/1016g** |
| **Cornstarch** | **397** | **397** | **334** | **387** | **114** | **378** |
| **Dextrinized cornstarch (90-94% tetrasaccharides)** | **132** | **131** | **110** | **127** | **37** | **124** |
| **Sucrose** | **100** | **100** | **84** | **97** | **28** | **95** |
| **Soybean oil** | **70** | **71** | **64** | **73** | **17** | **37** |
| **Cellulose fibers** | **50** | **50** | **0** | **50** | **0** | **50** |
| **Casein (≥85% protein)** | **200** | | | | | |
| **Soy protein isolate** | | **204** | | | | |
| **Spirulina powder** | | | **313** | | | |
| **Chickpea protein isolate** | | | | **210** | | |
| **Chickpea flour** | | | | | **841** | |
| **Fly larvae protein powder** | | | | | | **284** |
| **Mineral mix (AIN-93G-MX)** | **35** | **35** | **35** | **35** | **35** | **35** |
| **Vitamin mix (AIN-93G-MX)** | **10** | **10** | **10** | **10** | **10** | **10** |
| **Choline bitartrate (41.1% choline)** | **2.5** | **2.5** | **2.5** | **2.5** | **2.5** | **2.5** |
| **Tert-Butylhydroquinone** | **0.014** | **0.014** | **0.014** | **0.014** | **0.014** | **0.014** |
| **Nutritional level** | | | | | | |
| **Energy, kcal/g** | **3.2** | **2.87** | **3.42** | **3.14** | **3.2** | **2.882** |
| **Protein, %** | **21** | **21** | **21** | **21** | **21** | **21** |
| **Carbohydrate, %** | **61** | **61** | **61** | **61** | **61** | **61** |
| **Fat, %** | **18** | **18** | **18** | **18** | **18** | **18** |
| **Fibers, g/kg** | **50** | **50** | **62.7** | **50** | **90.9** | **50** |

**Bone structure analysis by computed microtomography (Micro-CT)**

Femurs were scanned with a Skyscan 1174 X-ray computed microtomography scanner device. Images were obtained by 50 kV X-ray tube voltages and 800 µA current, using a 0.25 mm aluminum filter, at 4000 ms exposure time, and a high spatial resolution of 15 µm. For each sample, a series of 900 projection images was obtained with a rotation step of 0.4°, averaging 2 frames, for a total of 360° rotation. A stack of 2D X-ray shadow projections was reconstructed to obtain images using NRecon software (Skyscan). Following reconstruction, images were subjected to morphometric analysis using CTAn software (Skyscan). Additional morphometric parameters were calculated as suggested by guidelines for bone microstructure assessment using micro-CT[28,29]. For the diaphyseal cortical region analysis, 200 slices, equivalent to 2.764 mm of the bone, were chosen. Global grayscale threshold levels for the cortical region were between 66-255. For the trabecular region, a total of 100 slices, equivalent to 1.382 mm of the bone, were selected, and adaptive grayscale threshold levels between 49 and 255 were used. Reconstructed scans were volume rendered (Amira software v.6.4, FEI, Hillsboro, OR, USA) to visualize the 2D and 3D morphology of the selected samples. Trabecular bone data included bone volume fraction (BV/TV), trabecular number (Tb.N), trabecular thickness (Tb.Th), and trabecular separation (Tb.Sp). For the cortical bone, data consisted of total cross-sectional area (Tt.Ar), cortical bone area (Ct.Ar), cortical area fraction (Ct.Ar/Tt.Ar), average cortical thickness (Ct.Th), medullary area (Ma.Ar) and bone mineral density (BMD). Additionally, the length of each femur bone was measured using the SkyScan software[26].

**Bone mechanical analysis by three-point bending**

For the three-point bending experiment, the right femora from each group were tested using an Instron material testing machine (Model 3345) fitted with custom-built saline containing a testing chamber as described before[30]. The distance between the stationary supports was set to 12 mm to ensure that the relatively tubular portion of the mid-diaphysis rests on these supports. An initial preload of 0.1 N was applied to hold the bone in place; Force-displacement data were collected by Instron software (BlueHill) at 10 Hz. The resulting force-displacement curves were used to calculate bone stiffness, bone yield point, load of fracture, maximal load, and area under the curve (AUC) to calculate the total energy to fracture. The slope of the linear portion of the curve was used to calculate bone stiffness that is affected both by bone geometry and material properties. Also, to determine the stiffness provided only by the material composition, Young's modulus was calculated. Young's modulus is the slope of the linear portion of a stress-strain curve, and it represents the stiffness of the material excluding bone length and architecture[31,32].

**Growth plate histology**

GPs of the tibiae were examined by a histological technique of Safranin-O staining. Tibiae were fixed, decalcified, dehydrated, and embedded in paraffin[33]. Transverse tissue sections of 5µm were prepared with Leica microtome (Agentec, Israel). Sections were stained in a Weigert's iron hematoxylin solution and fast green solution[34]. The stained sections were viewed by the light microscopy eclipse E400 Nikon. Images were captured by a high-resolution camera (Olympus DP 71) controlled by Cell A software (Olympos).

**Amino acid composition analyses**

Protein products were sent to MILOUDA&MIGAL and analytical laboratories (Israel) for amino acid analysis. Proteins were hydrolyzed, and the separated amino acids were quantitatively determined by using ion-exchange chromatography with an automatic analyzer[33].

**Microbiome analysis**

On the day of sacrifice, the caeca from all 48 rats were collected and stored at -80°C. Defrosted caecum (0.25 gr) content was extracted using a QIAamp PowerFecal kit (Qiagen, Hilden, Germany) according to the manufacturer's instructions. The extracted DNA was stored at -80ºc until sequencing of the V4 region using 515F-806R primers in HyLabs, Rehovot, Israel. Samples were run on a Miseq (Illumina) machine with 30% PhiX using a MiSeq Reagent Kit v2 500PE. Bioinformatical analysis in R (v.4.2.0) was conducted using the DADA2 package (v.1.24.0) for read quality control, primer trimming, read truncation and merging, and chimera removal. The DECIPHER package (v.2.24.0) and the RDP dataset (v.18) were used to assign taxonomy. The Phyloseq (v 1.40.0), Vegan (v 2.6.4), and DESeq2 (v 1.36.0) packages were used to calculate alpha and beta diversity, and differential abundance of genera. Beta diversity and differential abundance were calculated once normalized for library size. Differential abundance analysis was conducted after aggregating on the genus level and excluding genera with sum of counts across all samples <50. Amplicon sequence variants (ASVs) not classified on the domain level and bacteria not classified on the phylum level were excluded. The raw data is available on SRA: http://www.ncbi.nlm.nih.gov/bioproject/1132790 .

**Statistical analysis**

All growth and bone data are expressed as mean ± SD. The significance of differences between groups was determined using JMP 16.0. Statistical Discovery Software (SAS Institute 2000) by one-way analysis of variance if not otherwise stated. Differences between groups were further evaluated by Tukey-Kramer HSD test if not stated otherwise, considered significant at P<0.05. For the microbiome analysis, Wilcoxon rank sum test with FDR adjustment to assess differences in alpha diversity. For beta diversity analysis, PERMANOVA and Holm's adjustment were considered significant with p.adj < 0.05, if homogeneity of dispersion between groups was given (Tukey post-hoc and Holm's p. adj > 0.05). Differential abundant bacteria were considered significant if log2fold >1 and FDR p.adj <0.05.

**Supplements**

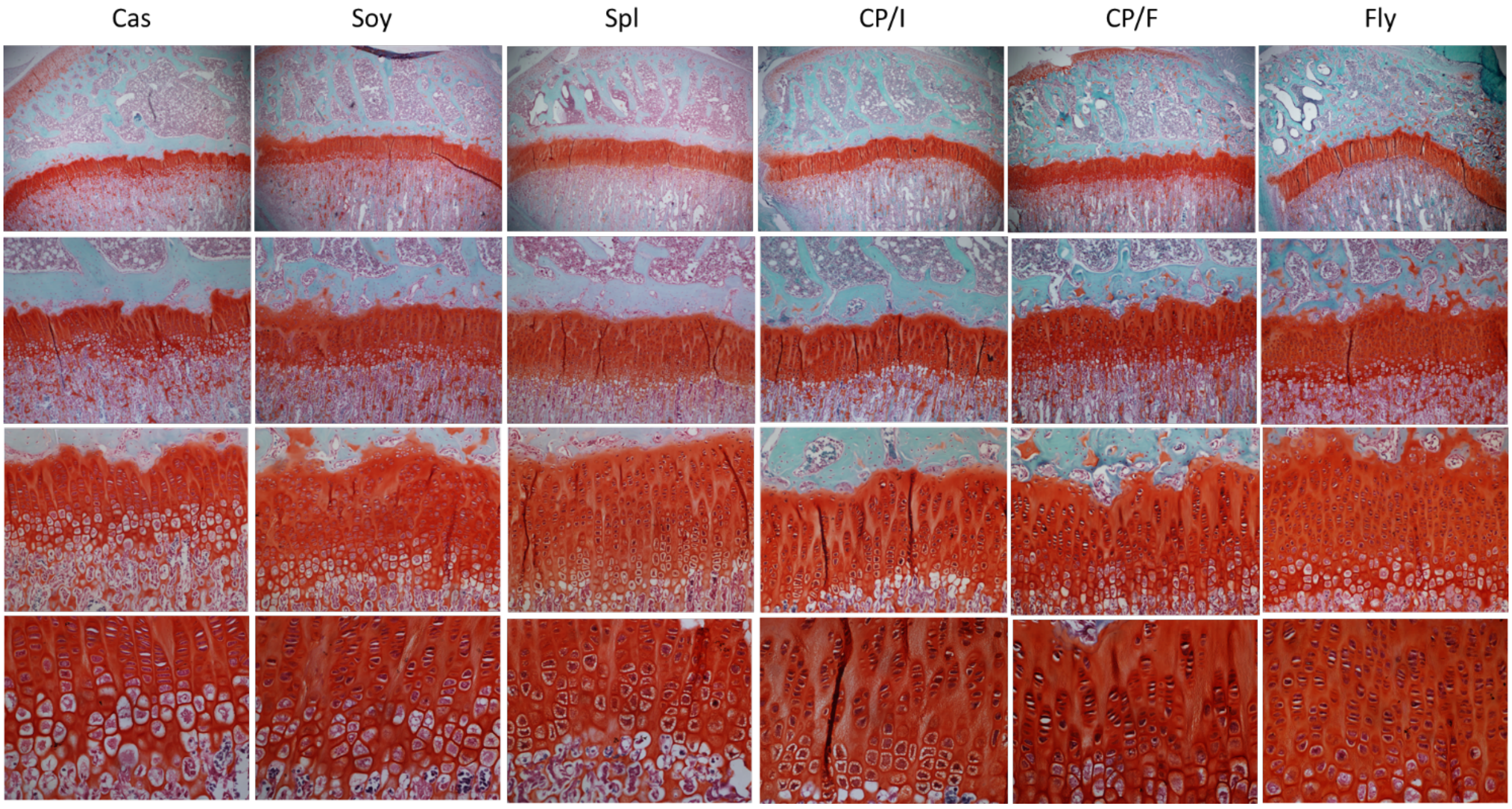

**Supplemental Figure 1. Histological evaluation of the tibial GP of 9-week-old rats**. Safranin-O staining of the tibial GP of 9-week-old rats from the 6 different diet groups.

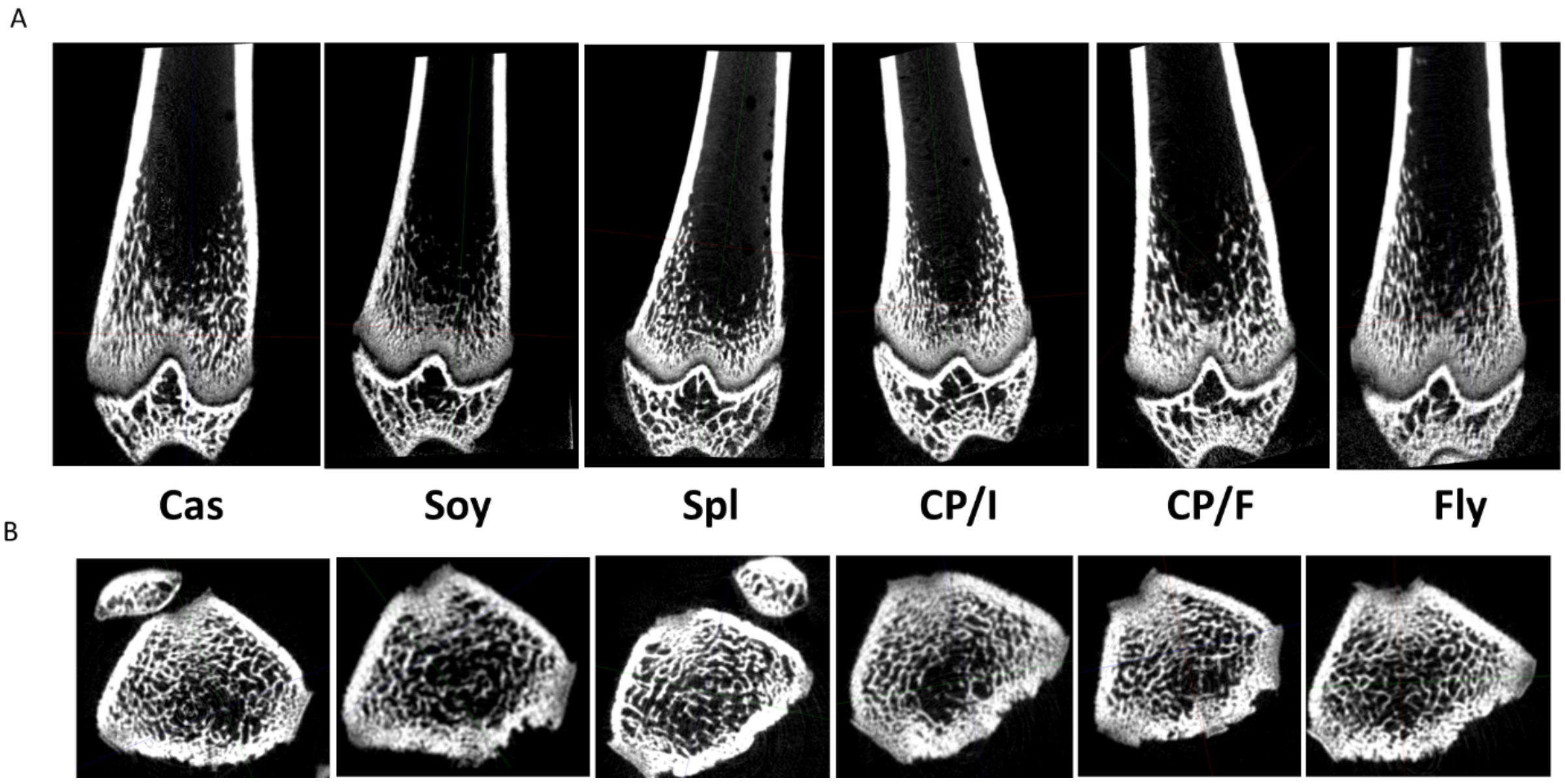

**Supplemental Figure 2. Bone architecture.** Three-dimensional reconstructions of femora scaled for mineral density using AMIRA software. A representative bone was taken from each of the six groups. (A) Sagittal sections of the bone. (B) Transverse sections of the trabecular area.

**Supplemental Table 1. Essential amino acid profiles of the tested protein sources, and the requirements for growing rats**. Values are expressed as g/kg diet. According to the NRC's amino acid requirements for growing rats

| g/kg diet (% of requirements) | Cas | Soy | Spl | CP/I | CP/F | Fly | Growing Rats Requirements |
|---|---|---|---|---|---|---|---|
| **Histidine** | 3.3(118%) | 4.4(157%) | 2.6(92%) | 3.8(136%) | 5.2(186%) | 4.1(146%) | 2.8 |
| **Isoleucine** | 5.3(85%) | 6.7(108%) | 6.6(106%) | 6.4(103%) | 6.6(106%) | 7.3(117%) | 6.2 |
| **Leucine** | 11.4(107%) | 12.8(120%) | 11.0(103%) | 12.4(116%) | 13.4(125%) | 12.8(120%) | 10.7 |
| **Lysine** | 11.8(128%) | 10.9(118%) | 7.4(80%) | 12.2(133%) | 13.4(146%) | 12.6(137%) | 9.2 |
| **Methionine+ Cystein** | 6.8(69%) | 3.9(40%) | 4.6(47%) | 3.4(35%) | 5.1(52%) | 4.5(46%) | 9.8 |
| **Phenylalanine+ Tyrosine** | 13.4(131%) | 15.4(151%) | 11.6(114%) | 14.4(141%) | 16.2(159%) | 18.8(184%) | 10.2 |
| **Threonine** | 5.4(87%) | 7.2(116%) | 7.0(113%) | 5.4(87%) | 8.1(131%) | 7.7(124%) | 6.2 |
| **Valine** | 7.3(99%) | 7.5(101%) | 6.4(86%) | 6.6(89%) | 7.6(103%) | 8.1(109%) | 7.4 |
| **Tryptophan** | 1.1(55%) | 1.76(88%) | 1.6(80%) | 1.0(50%) | 2.17(108%) | 1.8(90%) | 2 |

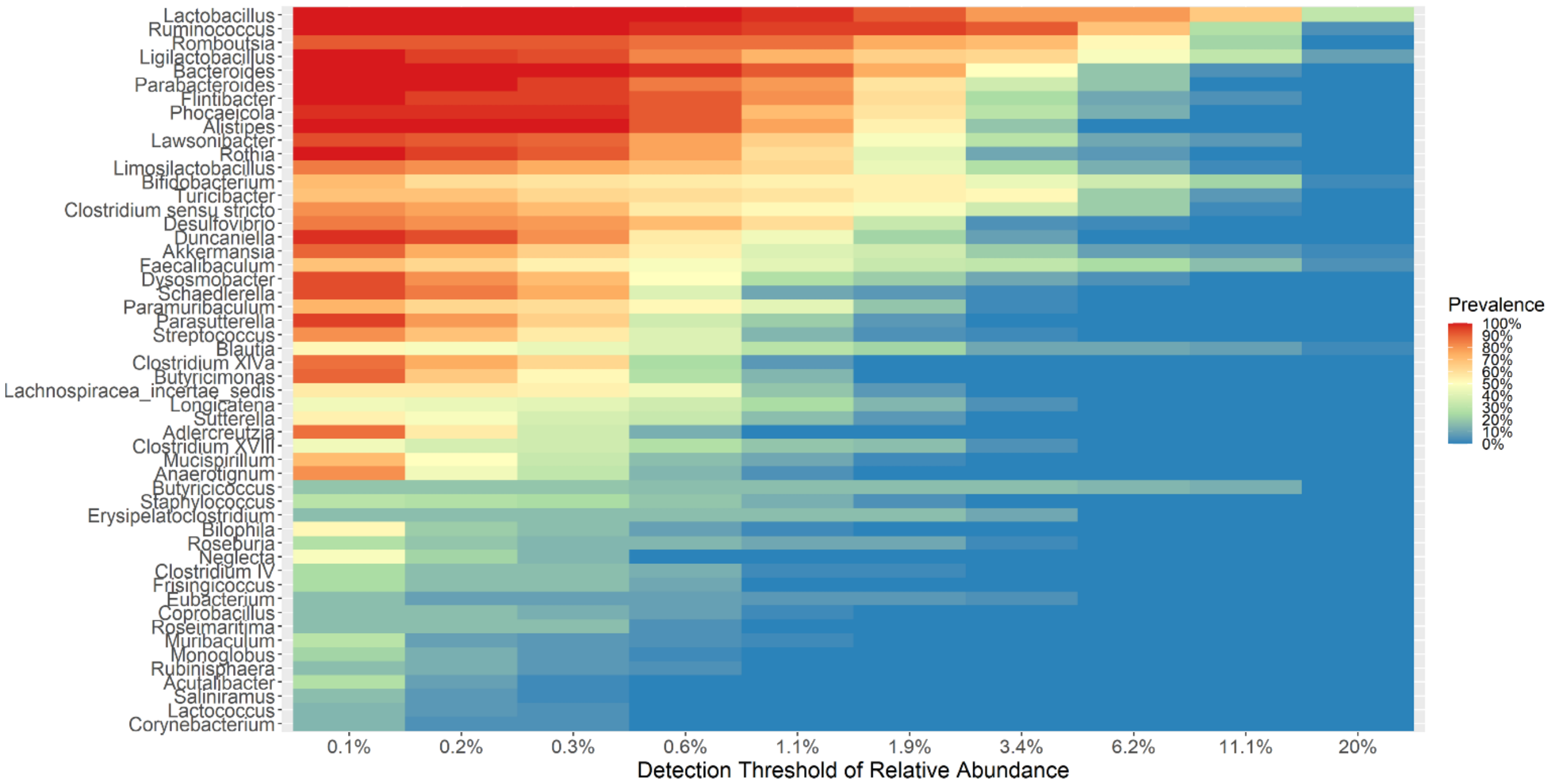

**Supplemental Figure 3**: Core microbiome. The prevalence of detected genera in relation to their relative abundance. Bacteria not classified at the genus level or below the minimum relative abundance threshold level were excluded.

**Supplemental Table 2:** Differential abundant genera comparing the control and the alternative protein groups

**Supplemental Table 3:** Differential abundant families comparing CP/I vs CP/F

**Supplemental Table 4:** Differential abundant genera comparing CP/I vs CP/F

| ASV ID | Control vs | log2FoldChange | lfcSE | p.adj | phylum | class | order |
|---|---|---|---|---|---|---|---|
| ASV2 | G_SOY_20 | -3.5 | 0.55 | 6.20E-07 | Firmicutes | Bacilli | Lactob |
| ASV3 | G_SOY_20 | -3.33 | 0.56 | 2.40E-06 | Firmicutes | Clostridia | Clostri |
| ASV6 | G_SOY_20 | -6.46 | 0.38 | 1.10E-61 | Firmicutes | Erysipelotrichia | Erysipe |
| ASV8 | G_SOY_20 | -4.14 | 0.65 | 2.20E-07 | Actinobacteria | Actinobacteria | Bifidob |
| ASV34 | G_SOY_20 | -4.44 | 0.88 | 1.20E-03 | Firmicutes | Bacilli | Lactob |
| ASV45 | G_SOY_20 | -1.19 | 0.25 | 2.30E-03 | Bacteroidetes | Bacteroidia | Bacter |
| ASV47 | G_SOY_20 | -4.9 | 1.09 | 8.10E-03 | Firmicutes | Clostridia | Clostri |
| ASV49 | G_SOY_20 | -3.52 | 0.49 | 2.00E-09 | Firmicutes | Erysipelotrichia | Erysipe |
| ASV79 | G_SOY_20 | 2.22 | 0.27 | 2.00E-12 | Bacteroidetes | Bacteroidia | Bacter |
| ASV277 | G_SOY_20 | 6.37 | 1.04 | 1.60E-06 | Firmicutes | Bacilli | Bacilla |
| ASV294 | G_SOY_20 | 8.91 | 0.84 | 5.30E-22 | Firmicutes | Clostridia | Clostri |
| ASV3 | G_SPL_20 | -4.85 | 0.49 | 6.40E-20 | Firmicutes | Clostridia | Clostri |
| ASV6 | G_SPL_20 | -8.71 | 0.37 | 5.20E-116 | Firmicutes | Erysipelotrichia | Erysipe |
| ASV8 | G_SPL_20 | -7.38 | 0.28 | 4.00E-145 | Actinobacteria | Actinobacteria | Bifidob |
| ASV10 | G_SPL_20 | 3.5 | 0.54 | 1.80E-07 | Firmicutes | Bacilli | Lactob |
| ASV12 | G_SPL_20 | -6.85 | 0.57 | 1.10E-29 | Firmicutes | Erysipelotrichia | Erysipe |
| ASV30 | G_SPL_20 | 3.26 | 0.37 | 3.00E-15 | Bacteroidetes | Bacteroidia | Bacter |
| ASV49 | G_SPL_20 | -7.4 | 0.78 | 1.20E-17 | Firmicutes | Erysipelotrichia | Erysipe |
| ASV75 | G_SPL_20 | 8.63 | 0.7 | 4.40E-31 | Firmicutes | Erysipelotrichia | Erysipe |
| ASV79 | G_SPL_20 | 2.56 | 0.33 | 4.90E-12 | Bacteroidetes | Bacteroidia | Bacter |
| ASV92 | G_SPL_20 | -8.99 | 0.86 | 4.90E-22 | Firmicutes | Clostridia | Clostri |
| ASV98 | G_SPL_20 | -2.48 | 0.41 | 2.50E-06 | Firmicutes | Clostridia | Clostri |
| ASV102 | G_SPL_20 | -1.15 | 0.28 | 3.80E-02 | Firmicutes | Clostridia | Clostri |
| ASV105 | G_SPL_20 | 2.13 | 0.49 | 1.30E-02 | Bacteroidetes | Bacteroidia | Bacter |
| ASV108 | G_SPL_20 | 10.59 | 0.82 | 1.50E-34 | Firmicutes | Erysipelotrichia | Erysipe |
| ASV146 | G_SPL_20 | 2.29 | 0.42 | 2.40E-05 | Firmicutes | Bacilli | Lactob |
| ASV162 | G_SPL_20 | 3.78 | 0.67 | 2.40E-05 | Proteobacteria | Betaproteobacteria | Burkho |
| ASV167 | G_SPL_20 | 8.11 | 0.84 | 1.90E-18 | Firmicutes | Erysipelotrichia | Erysipe |
| ASV428 | G_SPL_20 | 8.54 | 0.91 | 1.90E-17 | Planctomycetes | Planctomycetacia | Pirellul |
| ASV501 | G_SPL_20 | 3.58 | 0.81 | 1.20E-02 | Firmicutes | Bacilli | Lactob |
| ASV612 | G_SPL_20 | 7.84 | 0.96 | 8.40E-13 | Planctomycetes | Planctomycetacia | Planct |
| ASV650 | G_SPL_20 | 7.03 | 1.05 | 4.30E-08 | Proteobacteria | Alphaproteobacteria | Rhizob |
| ASV792 | G_SPL_20 | 6.2 | 1.18 | 1.60E-04 | Firmicutes | Erysipelotrichia | Erysipe |
| ASV1 | G_CP/I_20 | -2.48 | 0.37 | 2.60E-07 | Verrucomicrobia | Verrucomicrobiae | Verruc |
| ASV2 | G_CP/I_20 | -2.8 | 0.57 | 1.40E-03 | Firmicutes | Bacilli | Lactob |
| ASV6 | G_CP/I_20 | -3.18 | 0.72 | 5.30E-03 | Firmicutes | Erysipelotrichia | Erysipe |
| ASV8 | G_CP/I_20 | -2.96 | 0.71 | 2.10E-02 | Actinobacteria | Actinobacteria | Bifidob |
| ASV45 | G_CP/I_20 | -2.18 | 0.37 | 1.50E-05 | Bacteroidetes | Bacteroidia | Bacter |
| ASV49 | G_CP/I_20 | -2.18 | 0.45 | 1.00E-03 | Firmicutes | Erysipelotrichia | Erysipe |
| ASV79 | G_CP/I_20 | 2.26 | 0.29 | 6.30E-11 | Bacteroidetes | Bacteroidia | Bacter |
| ASV294 | G_CP/I_20 | 4.85 | 1.08 | 1.00E-02 | Firmicutes | Clostridia | Clostri |
| ASV312 | G_CP/I_20 | 5.33 | 0.65 | 6.80E-12 | Firmicutes | Clostridia | Clostri |
| ASV1 | G_CP/F_20 | -2.57 | 0.51 | 4.20E-04 | Verrucomicrobia | Verrucomicrobiae | Verruc |
| ASV3 | G_CP/F_20 | -7.13 | 0.5 | 3.20E-41 | Firmicutes | Clostridia | Clostri |
| ASV4 | G_CP/F_20 | -3.34 | 0.75 | 1.20E-02 | Firmicutes | Clostridia | Clostri |
| ASV6 | G_CP/F_20 | -2.93 | 0.49 | 1.10E-06 | Firmicutes | Erysipelotrichia | Erysipe |
| ASV12 | G_CP/F_20 | -8.35 | 0.63 | 4.30E-36 | Firmicutes | Erysipelotrichia | Erysipe |
| ASV16 | G_CP/F_20 | -2.84 | 0.65 | 1.50E-02 | Firmicutes | Clostridia | Clostri |
| ASV23 | G_CP/F_20 | -5.6 | 0.68 | 4.90E-13 | Proteobacteria | Deltaproteobacteria | Desulf |
| ASV30 | G_CP/F_20 | 2.06 | 0.47 | 5.00E-03 | Bacteroidetes | Bacteroidia | Bacter |
| ASV45 | G_CP/F_20 | 1.98 | 0.38 | 2.20E-04 | Bacteroidetes | Bacteroidia | Bacter |
| ASV47 | G_CP/F_20 | 6.17 | 0.89 | 1.10E-08 | Firmicutes | Clostridia | Clostri |
| ASV49 | G_CP/F_20 | -7.79 | 0.97 | 1.80E-12 | Firmicutes | Erysipelotrichia | Erysipe |
| ASV59 | G_CP/F_20 | -6.37 | 1.37 | 5.20E-03 | Deferribacteres | Deferribacteres | Deferri |
| ASV61 | G_CP/F_20 | 13.14 | 0.81 | 5.20E-54 | Firmicutes | Clostridia | Clostri |
| ASV75 | G_CP/F_20 | 8.98 | 0.79 | 7.90E-27 | Firmicutes | Erysipelotrichia | Erysipe |
| ASV79 | G_CP/F_20 | 1.8 | 0.37 | 4.30E-04 | Bacteroidetes | Bacteroidia | Bacter |
| ASV92 | G_CP/F_20 | -8.57 | 0.92 | 3.20E-17 | Firmicutes | Clostridia | Clostri |
| ASV98 | G_CP/F_20 | -2.77 | 0.66 | 1.50E-02 | Firmicutes | Clostridia | Clostri |
| ASV122 | G_CP/F_20 | 2.77 | 0.37 | 2.00E-10 | Proteobacteria | Betaproteobacteria | Burkho |
| ASV287 | G_CP/F_20 | 9.66 | 0.91 | 9.10E-23 | Firmicutes | Clostridia | Clostri |
| ASV833 | G_CP/F_20 | 6.45 | 1.27 | 6.40E-04 | Firmicutes | Clostridia | Clostri |
| ASV1206 | G_CP/F_20 | 6.98 | 1.27 | 7.40E-05 | Firmicutes | Erysipelotrichia | Erysipe |
| ASV1 | G_FLY_20 | 1.67 | 0.41 | 2.80E-02 | Verrucomicrobia | Verrucomicrobiae | Verruc |
| ASV2 | G_FLY_20 | -1.14 | 0.2 | 1.80E-05 | Firmicutes | Bacilli | Lactob |
| ASV3 | G_FLY_20 | -2.66 | 0.63 | 1.40E-02 | Firmicutes | Clostridia | Clostri |
| ASV6 | G_FLY_20 | -9.22 | 0.55 | 2.30E-60 | Firmicutes | Erysipelotrichia | Erysipe |
| ASV8 | G_FLY_20 | -8.06 | 0.32 | 3.20E-140 | Actinobacteria | Actinobacteria | Bifidob |
| ASV10 | G_FLY_20 | 3.22 | 0.57 | 3.20E-05 | Firmicutes | Bacilli | Lactob |
| ASV30 | G_FLY_20 | 2.61 | 0.38 | 1.90E-08 | Bacteroidetes | Bacteroidia | Bacter |
| ASV47 | G_FLY_20 | -4.68 | 1.12 | 2.20E-02 | Firmicutes | Clostridia | Clostri |
| ASV49 | G_FLY_20 | 2.21 | 0.49 | 3.70E-03 | Firmicutes | Erysipelotrichia | Erysipe |
| ASV75 | G_FLY_20 | 5 | 1 | 6.10E-04 | Firmicutes | Erysipelotrichia | Erysipe |
| ASV146 | G_FLY_20 | 1.61 | 0.27 | 8.90E-06 | Firmicutes | Bacilli | Lactob |
| ASV277 | G_FLY_20 | 9.54 | 0.91 | 6.80E-22 | Firmicutes | Bacilli | Bacilla |

| ASV ID | log2FoldChange | lfcSE | p.adj | phylum | class | order | family |
|---|---|---|---|---|---|---|---|
| ASV2 | 1.04 | 0.46 | 3.36E-02 | Firmicutes | Bacilli | Lactobacillales | Lactobaci |
| ASV3 | -5.93 | 0.60 | 3.10E-22 | Firmicutes | Clostridia | Clostridiales | Clostridia |
| ASV4 | -4.57 | 0.96 | 6.76E-06 | Firmicutes | Clostridia | Clostridiales | Peptostre |
| ASV8 | 2.69 | 0.62 | 3.59E-05 | Actinobacteria | Actinobacteria | Bifidobacteriales | Bifidobact |
| ASV23 | -3.73 | 0.60 | 2.89E-09 | Proteobacteria | Deltaproteobacteria | Desulfovibrionales | Desulfovib |
| ASV26 | -1.56 | 0.46 | 1.20E-03 | Actinobacteria | Actinobacteria | Micrococcales | Micrococc |
| ASV30 | 1.48 | 0.40 | 5.15E-04 | Bacteroidetes | Bacteroidia | Bacteroidales | Porphyror |
| ASV38 | 1.15 | 0.39 | 5.65E-03 | Firmicutes | Clostridia | Clostridiales | Lachnosp |
| ASV58 | -2.08 | 0.35 | 1.43E-08 | Bacteroidetes | Bacteroidia | Bacteroidales | Muribacul |
| ASV59 | -7.75 | 0.82 | 1.97E-20 | Deferribacteres | Deferribacteres | Deferribacterales | Deferribac |
| ASV105 | 1.87 | 0.47 | 1.30E-04 | Bacteroidetes | Bacteroidia | Bacteroidales | Odoribact |
| ASV122 | 2.74 | 0.46 | 1.09E-08 | Proteobacteria | Betaproteobacteria | Burkholderiales | Sutterella |
| ASV146 | -3.34 | 0.49 | 8.30E-11 | Firmicutes | Bacilli | Lactobacillales | Streptoco |

| **ASV ID** | **log2FoldChange** | **lfcSE** | **p.adj** | **phylum** | **class** |
|---|---|---|---|---|---|
| ASV2 | 3.15 | 0.58 | 1.34E-07 | Firmicutes | Bacilli |
| ASV3 | -5.22 | 0.63 | 4.83E-16 | Firmicutes | Clostridia |
| ASV4 | -3.83 | 0.82 | 6.67E-06 | Firmicutes | Clostridia |
| ASV8 | 3.47 | 0.65 | 2.14E-07 | Actinobacteria | Actinoba |
| ASV10 | -2.59 | 0.65 | 1.21E-04 | Firmicutes | Bacilli |
| ASV12 | -7.45 | 0.58 | 7.55E-37 | Firmicutes | Erysipelo |
| ASV16 | -3.57 | 0.55 | 3.29E-10 | Firmicutes | Clostridia |
| ASV23 | -5.63 | 0.61 | 1.56E-19 | Proteobacteria | Deltaprot |
| ASV30 | 2.14 | 0.44 | 2.66E-06 | Bacteroidetes | Bacteroid |
| ASV45 | 4.19 | 0.41 | 1.69E-23 | Bacteroidetes | Bacteroid |
| ASV47 | 8.11 | 0.87 | 6.10E-20 | Firmicutes | Clostridia |
| ASV49 | -5.72 | 0.97 | 1.01E-08 | Firmicutes | Erysipelo |
| ASV52 | 1.60 | 0.48 | 1.50E-03 | Bacteroidetes | Bacteroid |
| ASV59 | -7.15 | 0.86 | 6.07E-16 | Deferribacteres | Deferriba |
| ASV61 | 13.01 | 0.80 | 3.98E-57 | Firmicutes | Clostridia |
| ASV75 | 10.48 | 0.86 | 2.99E-33 | Firmicutes | Erysipelo |
| ASV81 | 1.90 | 0.71 | 1.02E-02 | Proteobacteria | Deltaprot |
| ASV87 | -3.73 | 0.90 | 6.08E-05 | Bacteroidetes | Bacteroid |
| ASV92 | -7.64 | 0.91 | 2.75E-16 | Firmicutes | Clostridia |
| ASV98 | -2.27 | 0.77 | 4.40E-03 | Firmicutes | Clostridia |
| ASV102 | -1.93 | 0.35 | 1.13E-07 | Firmicutes | Clostridia |
| ASV105 | 2.52 | 0.48 | 5.07E-07 | Bacteroidetes | Bacteroid |
| ASV122 | 3.58 | 0.52 | 3.54E-11 | Proteobacteria | Betaprote |
| ASV146 | -2.69 | 0.56 | 3.97E-06 | Firmicutes | Bacilli |
| ASV152 | -3.16 | 0.65 | 2.70E-06 | Firmicutes | Clostridia |
| ASV169 | -1.27 | 0.41 | 3.23E-03 | Firmicutes | Clostridia |
| ASV287 | 9.88 | 0.89 | 2.12E-27 | Firmicutes | Clostridia |
| ASV293 | -4.59 | 1.95 | 2.55E-02 | Firmicutes | Clostridia |
| ASV294 | -3.67 | 0.99 | 3.83E-04 | Firmicutes | Clostridia |
| ASV312 | -3.99 | 0.50 | 4.97E-15 | Firmicutes | Clostridia |
| ASV395 | -3.66 | 0.83 | 2.26E-05 | Firmicutes | Clostridia |
| ASV712 | -3.33 | 1.35 | 1.93E-02 | Firmicutes | Clostridia |
| ASV717 | -3.98 | 1.30 | 3.58E-03 | Firmicutes | Clostridia |
| ASV833 | 4.98 | 0.88 | 4.36E-08 | Firmicutes | Clostridia |
| ASV1206 | 7.18 | 1.01 | 4.59E-12 | Firmicutes | Erysipelo |